\documentclass[12pt]{article}
\usepackage{graphicx,amsmath}
\usepackage{graphicx,amsmath,lineno,color}
\newcommand{\cred}[1]{{\color{black} #1}}
\usepackage{units}

\newlength{\dinwidth}
\newlength{\dinmargin}
\def\lapproxeq{\lower .7ex\hbox{$\;\stackrel{\textstyle                                                    
<}{\sim}\;$}}                                                    
\def\gapproxeq{\lower .7ex\hbox{$\;\stackrel{\textstyle                                                    
>}{\sim}\;$}}                                                    
\def\be{\begin{equation}}                                                    
\def\ee{\end{equation}}                                                    
\def\bea{\begin{eqnarray}}                                                    
\def\eea{\end{eqnarray}}

\def\sh{\hat s}
\def\sh2{{\hat s}^2}

\begin{document}
                                                    
\titlepage                                                    
\begin{flushright}                                                    
IPPP/19/79  \\                                                    
\today \\                                                    
\end{flushright} 
\vspace*{0.5cm}

\begin{center}                                                    
{\Large \bf Bethe phase including proton excitations}\\
\vspace*{1cm}
                                                   
V.A. Khoze$^{a,b}$, A.D. Martin$^a$ and M.G. Ryskin$^{a,b}$ \\\                                                    
                                                   
\vspace*{0.5cm}                                                    
$^a$ Institute for Particle Physics Phenomenology, University of Durham, Durham, DH1 3LE \\                                                   
$^b$ Petersburg Nuclear Physics Institute, NRC Kurchatov Institute, Gatchina, St.~Petersburg, 188300, Russia

\vspace*{1cm}

\begin{abstract}
We evaluate the contribution of inelastic intermediate states ({\cred {such as}} $p\to N^*$ excitations) to the phase between the one-photon-exchange and the `nuclear' high energy 
{\cred {$pp$ scattering}}
amplitudes as $t\to 0$, caused by the multiphoton diagrams. It turns out to be rather small - much smaller than to have any influence on the experimental accuracy of {\cred {the measurements 
}}of $\rho$, defined to be  the ratio of the real to imaginary parts of the forward `nuclear' amplitude.
\end{abstract}

\end{center}

\vspace{1cm}

\section{Introduction}
The conventional way to measure the real part of the strong interaction (nuclear)
forward amplitude, $F^N$, is to consider its interference with the pure real one-photon-exchange QED amplitude, $F^C$, at very small momentum transfer $t\to 0$. However this interference is affected by the possibility of  multiphoton exchange processes which 
{\cred {result}} 
in additional phase difference $\alpha\phi$.
That is, the  total amplitude reads
\begin{equation}
\label{1}
F^{TOT}~=~F^N~+~e^{i\alpha\phi}F^C\ .
\end{equation}
Here $\alpha=\alpha^{\rm QED}=1/137$. The phase $\phi$ {\cred{(the so-called
Bethe phase)}} was calculated first by Bethe~\cite{Be} using {\cred {a}}
 WKB approach, and then was re-examined by West and Yennie \cite{WY} in terms of Feynman diagrams. A more precise calculation, which accounts for the details of the proton form factor, was performed by Cahn \cite{Cahn}.  It gives
\begin{equation}
\label{2}
\phi(t)~=~-[\ln(-Bt/2)+\gamma_E+C]\ ,
\end{equation}
where $B$ is the $t$-slope of the elastic cross section
{\cred { ($d\sigma_{\rm el}/dt\propto e^{Bt})$,}} $\gamma_E=0.577...$ is Euler's constant and 
{\cred{ $C=0.62$ (0.60)}}  depending on 
{\cred {which form of the proton}} electromagnetic form factor - exponential (or dipole)~\footnote{Note that the value of C=0.62 (0.60) was calculated in
 ~\cite{Cahn} for the ISR energies,
 assuming $B=13$ GeV$^{-2}$. In the LHC
 case with $B=20$ GeV$^{-2}$ we get C=0.45.}
{\cred {is}} used (see also \cite{SL1} for a more detailed calculation).

Note that in all  previous calculations only the pure eikonal diagrams were considered. That is only the `elastic' ($p\to p$) intermediate states were allowed
in the multiphoton exchange diagrams Fig.1a,b.~\footnote{Actually working at $O(\alpha)$ accuracy it is sufficient to study only the 
two-photon exchange QED diagram and one additional photon in the nuclear amplitude.} Besides this, there are {\cred {diagrams with the proton}}
excitations shown in Fig.1c,d. Of course, at small $t$ due to gauge invariance the $p+\gamma\to N^*$ vertex {\cred{contains transverse momentum}} $q_{t\gamma}$.
 Therefore, these diagrams do not {\cred {generate}} $\ln|t|$ and can only affect the value of the constant $C$. 

In the case of the TOTEM experiment at $\sqrt s=13$ TeV the value of $\rho$ was extracted by fitting the differential $d\sigma_{\rm el}/dt$ proton-proton elastic 
cross section in the region of very small $|t|\sim 0.001-0.005$ GeV$^2$,  where the role of the Bethe phase is not negligible. It changes the resulting value of $\rho \equiv$ Re/Im ratio by about 0.03. This should be compared with the experimental accuracy 0.01 ($\rho=0.10\pm 0.01$~\cite{TOTEM}). 
 However, the variation of $C$ by $\delta C=O(1)$ may additionally shift the value of $\rho$ by $\delta\rho=0.01~-~0.02$. Such {\cred {an effect could potentially be}}  important for the confirmation of the possible presence of the odd signature (Odderon) contribution in {\cred {the}} high energy $pp$-amplitude at $t\to 0$.
 Indeed the value {\cred {of}} $\rho=0.10\pm 0.01$, extracted using the phase $\phi$ calculated in \cite{Cahn}
 (without accounting for the possibility of proton excitation)
  is noticeably lower than that ($\rho\simeq 0.135$) {\cred{obtained}} from dispersion relations for a pure even-signature amplitude (with the total cross sections measured by TOTEM).
 The observed difference
$0.135-0.10=0.035\pm 0.01$ can be explained either by the odd-signature nuclear contribution to elastic $pp$ scattering or by {\cred {a}} modification of the constant $C$
due to the diagrams {\cred {of Figs.1c,d}} with inelastic ($p\to N^*$) intermediate states.

Therefore, it is timely to evaluate the possible role of {\cred {the}} processes with proton 
 excitations in {\cred {the}} Coulomb-nuclear interference region.
Unfortunately, there are no sufficient data on diffractive  $p\to N^*$ dissociation which would allow the calculation of the {\cred {contribution of Fig.1d}} explicitly.  On the other hand, it is known that cross section of low-mass diffractive excitation is well described 
by the so-called Deck $p\to N+\pi$ process~\cite{Deck}, shown in Fig.2a.   

Therefore, in section 2 we {\cred {use}} 
the diagrams of Fig.2a to evaluate the expected
shift $\delta C$ caused by low-mass {\cred {excitations}}. The higher-mass contribution 
{\cred {is}} calculated in section 3 {\cred {based}} on the triple-Regge formalism 
{\cred{(Fig.2b) 
and duality}}. Next, in section 4, we calculate the phase shift {\color{black}$\delta\phi^C$} originating from the two-photon graph Fig.1c. Here data on {\cred {the}} $\gamma p$ cross sections are {\cred {available}} and will be used. Unlike one-photon exchange this diagram does not {\cred {contain}} a factor of $1/t$. Thus, at very small $t\to 0$ the corresponding correction is strongly suppressed and can be neglected.
Besides this, formally the diagram {\cred{in}} Fig.1c describes the even-signature amplitude
 and should satisfy even-signature dispersion relations.
  We conclude in section 5.
\begin{figure}[hbt]\vspace{-1cm}
\vspace{-2cm}
\begin{center}
  \includegraphics[scale=0.35]{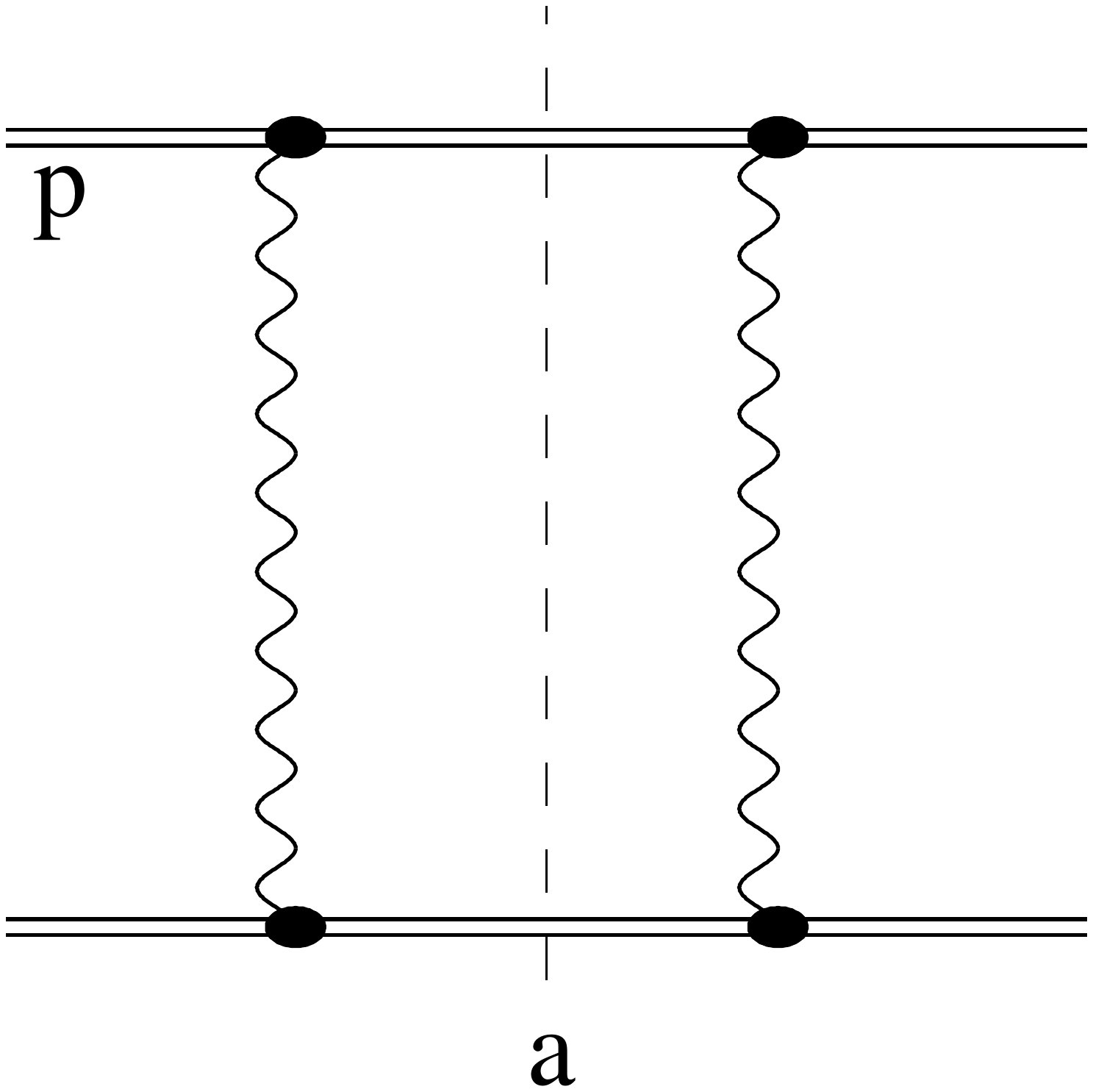}
\includegraphics[scale=0.35]{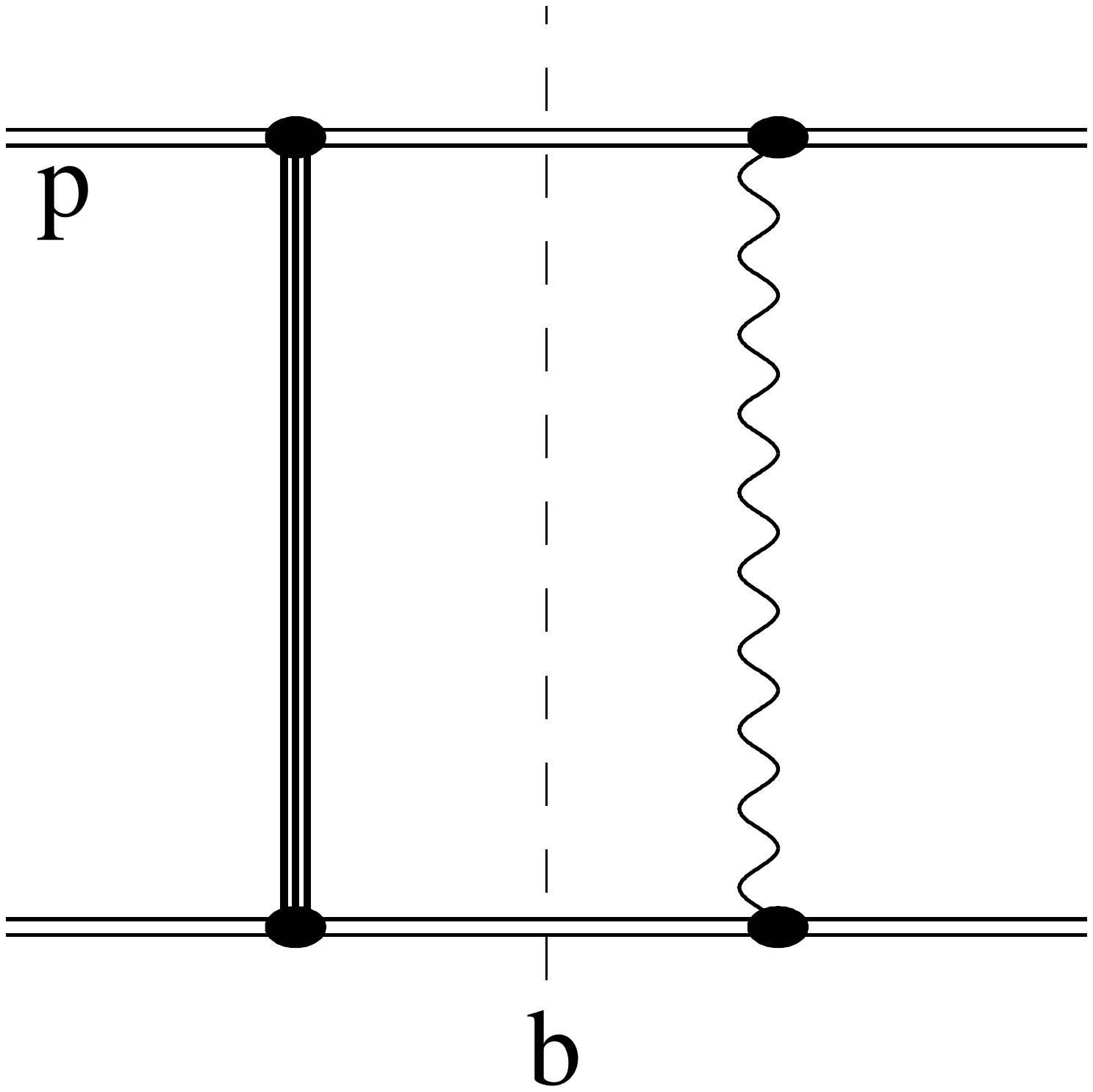}\vspace{-3cm}
\vspace{1cm}
\label{fig:0}
\end{center} 
\vspace{-2cm}
\hspace{0.7cm}
\includegraphics[scale=0.35]{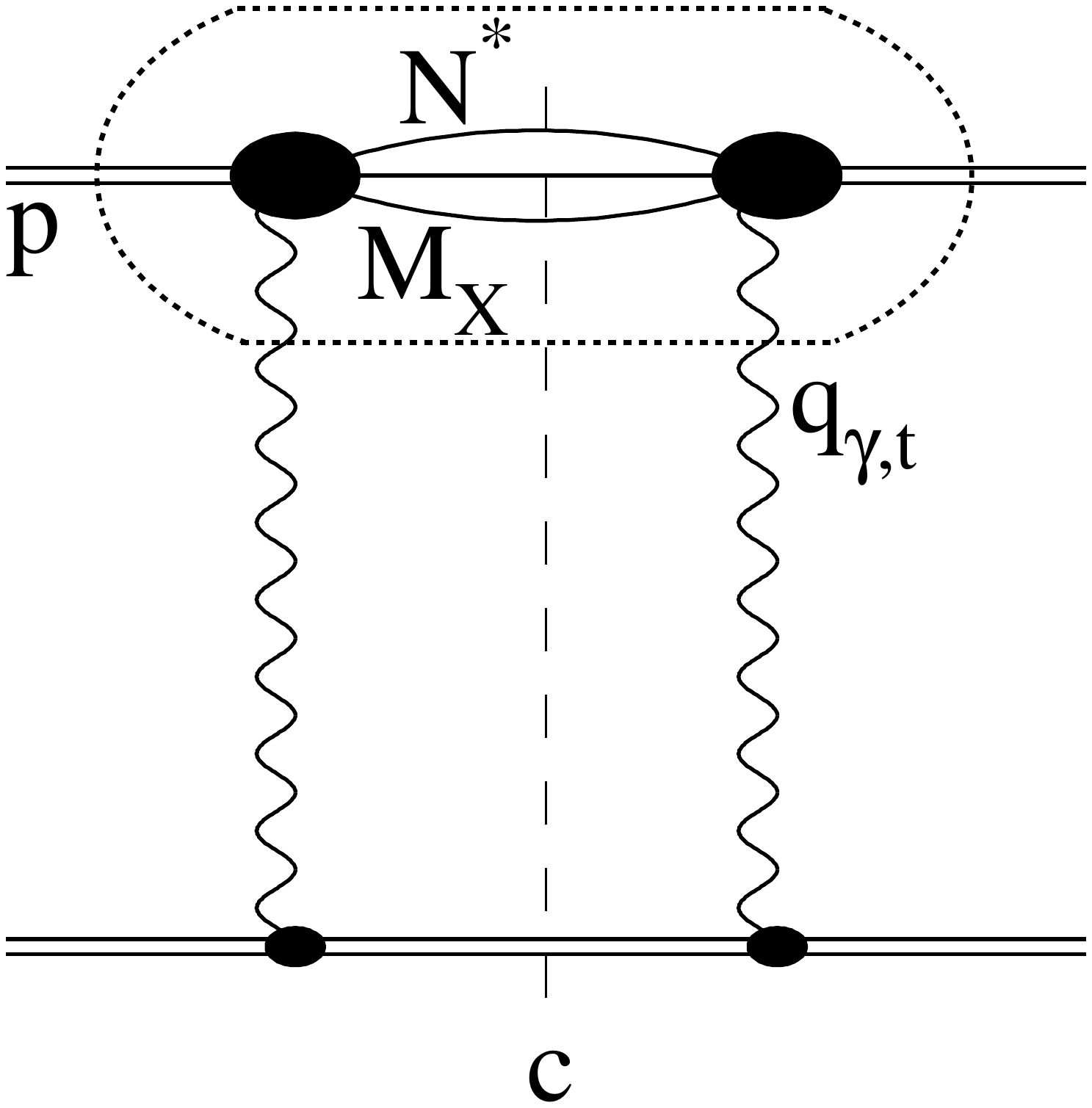}
\includegraphics[scale=0.35]{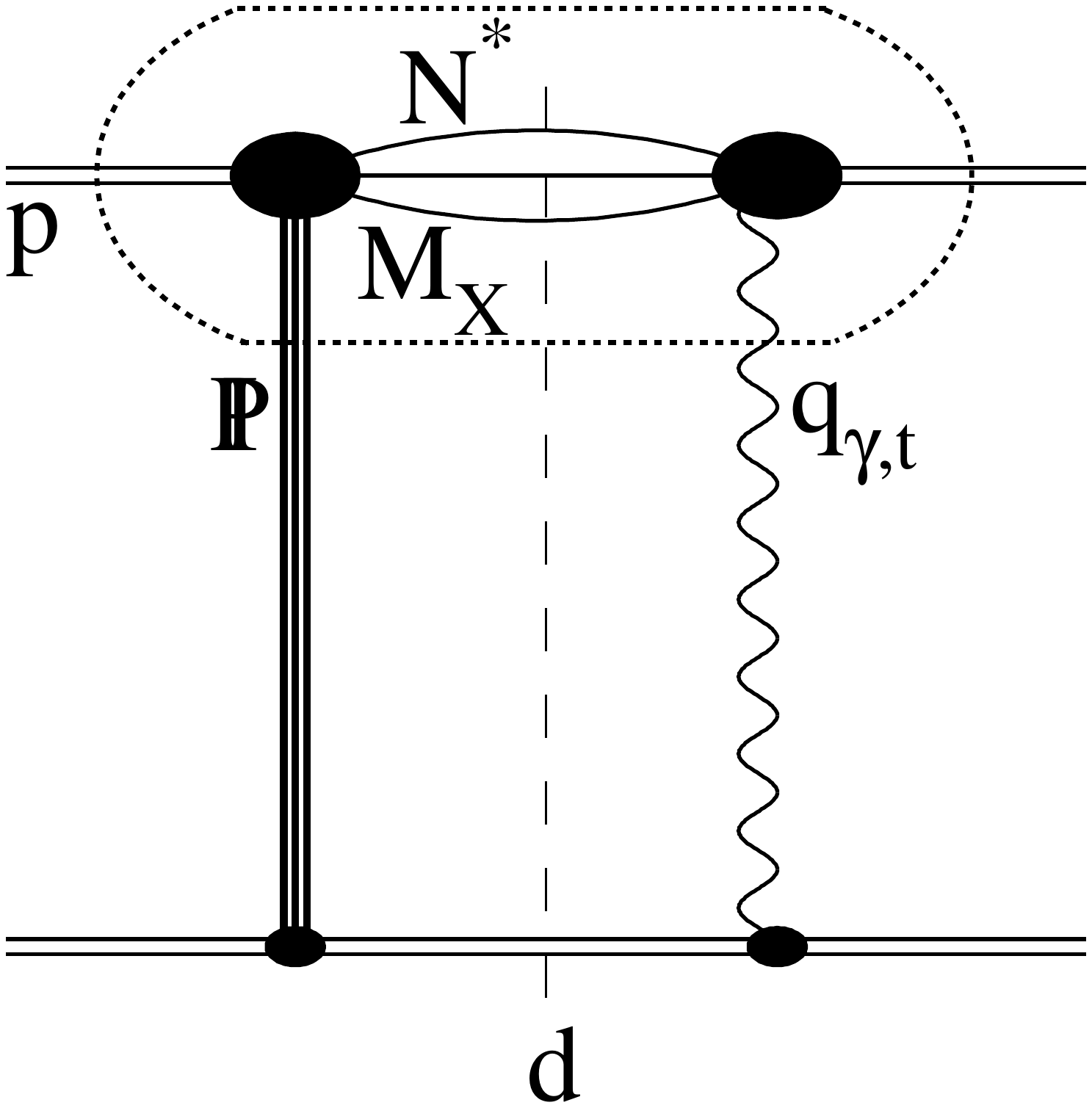}
\caption{\sf Diagrams responsible for the Bethe phase at the lowest $\alpha^{\rm QED}$  order. The four plots are: a- the eikonal (elastic) phase of the one-photon-exchange amplitude, b- the `elastic' phase of the strong interaction amplitude, c- and d- are the contributions of the excited ($p\to N^*$) intermediate states. The nuclear amplitude is shown by the triple  solid  line and marked as I\!P.}
\label{fig:1}
\end{figure}

\section{Phase shift caused by {\cred {the}} Deck process}
At the lowest $\alpha^{\rm QED}$ order the phase of the strong interaction amplitude (marked in figures as $I\!\!P$) is given by the {\em discontinuity} shown in {\cred {Figs.1b,d}} by the vertical dashed lines. Taking the discontinuity of the amplitude, that is replacing $i\pi$ by $2i\pi$ in the imaginary part of 
{\cred {the}} propagator we account for the contribution where the photon exchange is now placed to the left of the nuclear amplitude. Besides this, in  
{\cred {Fig.1d (and also in Figs 3 and 4)}} we have to include {\cred {an additional}} factor of $2$ since the lower proton
{\cred {can also}}
 dissociate.
 \begin{center}
 \begin{figure}[htb]
 \vspace{-3cm}
\hspace{1.6cm}
\includegraphics[scale=0.35]{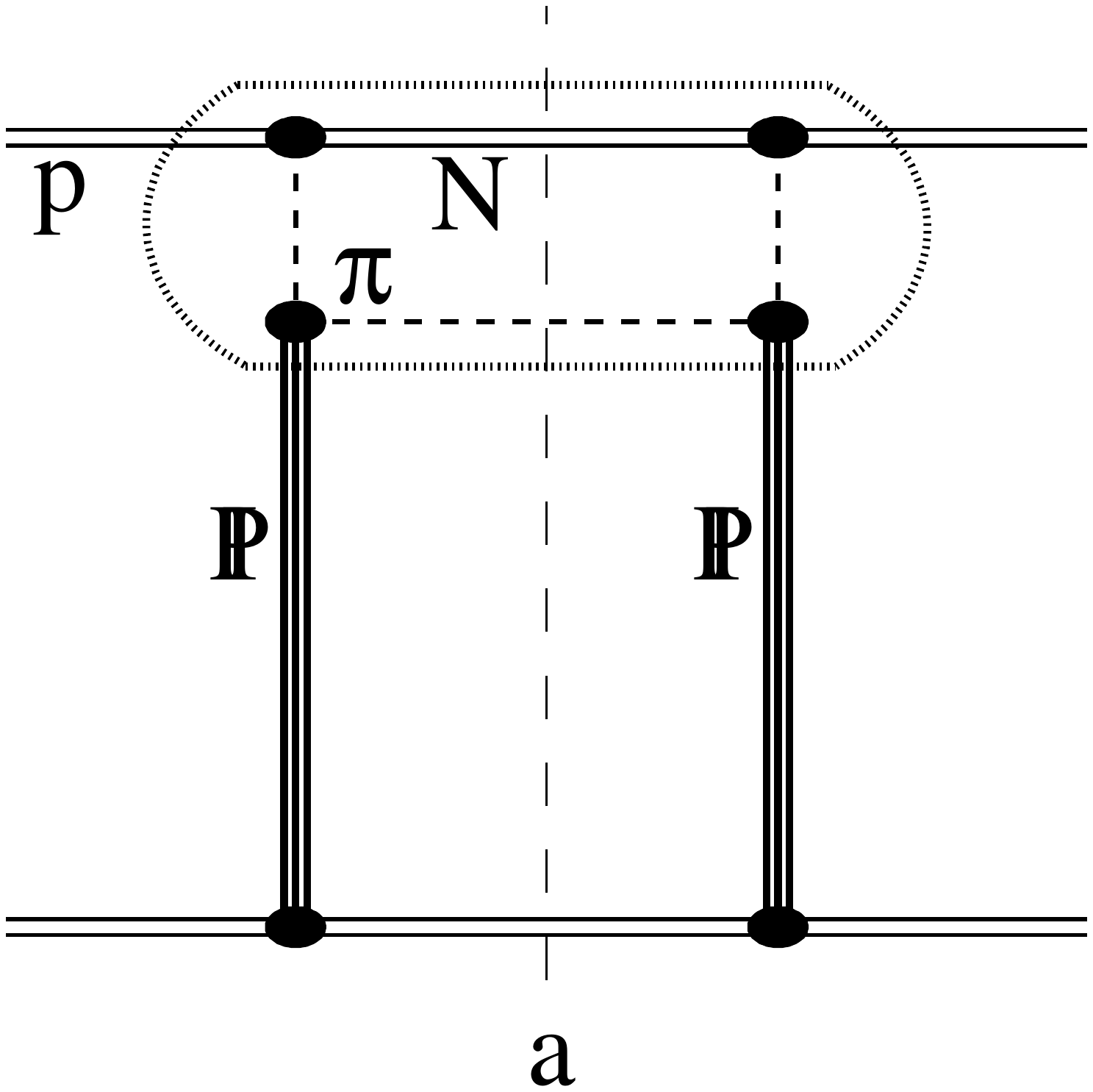}
\hspace{-1cm}
\includegraphics[scale=0.35]{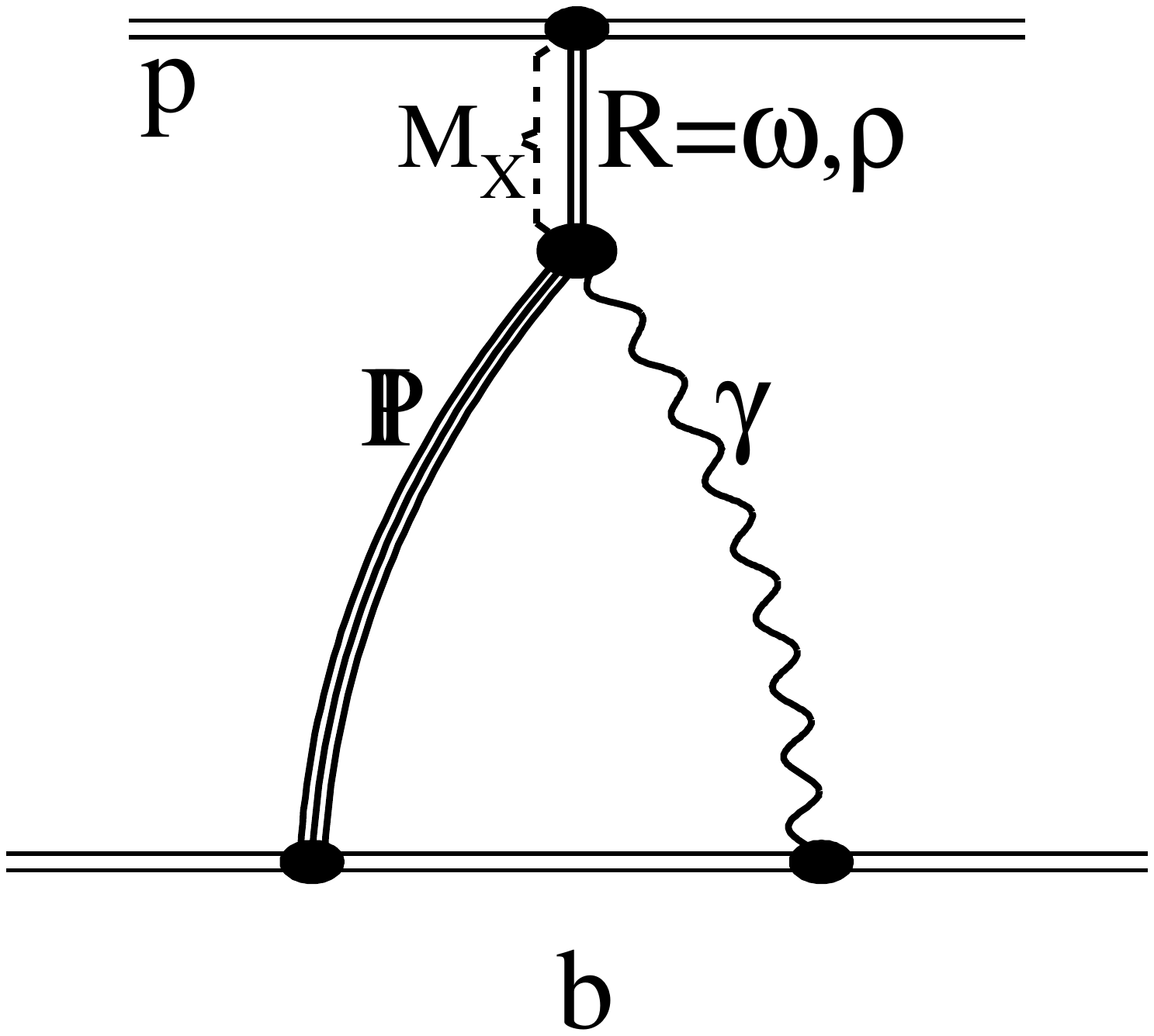}
\vspace{-1cm}
\caption{\sf a- The Deck diagram for the low-mass proton dissociation. b- The diagram of triple-Regge form used to evaluate, via {\cred {duality}, } the contribution of {\cred {the}} heavier intermediate states.}
\label{fig:2}
\end{figure}
\end{center}
Instead of the $p\to N^*$ low-mass excitation we consider the simplest diagrams for the $p\to N\pi$ transition which rather well reproduce the low-mass proton dissociation~\cite{Deck}. In particular, the cross section of diffractive dissociation calculated via
{\cred {the diagram of Fig.2a}}  at {\cred {the}} LHC energy $\sqrt s=7$ GeV is about 2.7 mb (for dissociation of both - that is either the upper or lower - protons). This is close to the value of low-mass dissociation ($\sigma^{\rm SD}(M_X<3.4{\rm GeV})=2.66\pm 2.17$mb) measured by TOTEM~\cite{TOTEM2} (see also the discussion in sect.3 
of~\cite{KMR18}).

Recall that the amplitude of {\cred {the}} Deck processes is described by {\em three} diagrams shown in Fig.3. For the photon exchange amplitude we have an analogous set of three diagrams (Fig.4). That is, to calculate the discontinuity we have to sum up the three diagrams of Fig.3 and multiply {\cred {this contribution}}
 by the sum of {\cred {the three diagrams of Fig.4 }}.

Since we are looking just for {\cred {an}} additional Bethe phase which may affect the $\rho={\rm Re/Im}$ ratio we {\cred{should not  worry}} about {\cred{an}} exact value of {\cred {the}} high energy strong amplitude. We use the normalization $s\sigma_{\rm tot}={\rm Im}T(t=0)$  assuming that $\sigma_{\rm tot}\propto s^{0.1}$ and use the additive quark model relation $\sigma(\pi p)=(2/3)\sigma(pp)$. 

  
 \begin{figure}[htb]
\vspace{-3cm}
\hspace{1.5cm} 
\includegraphics[scale=0.33]{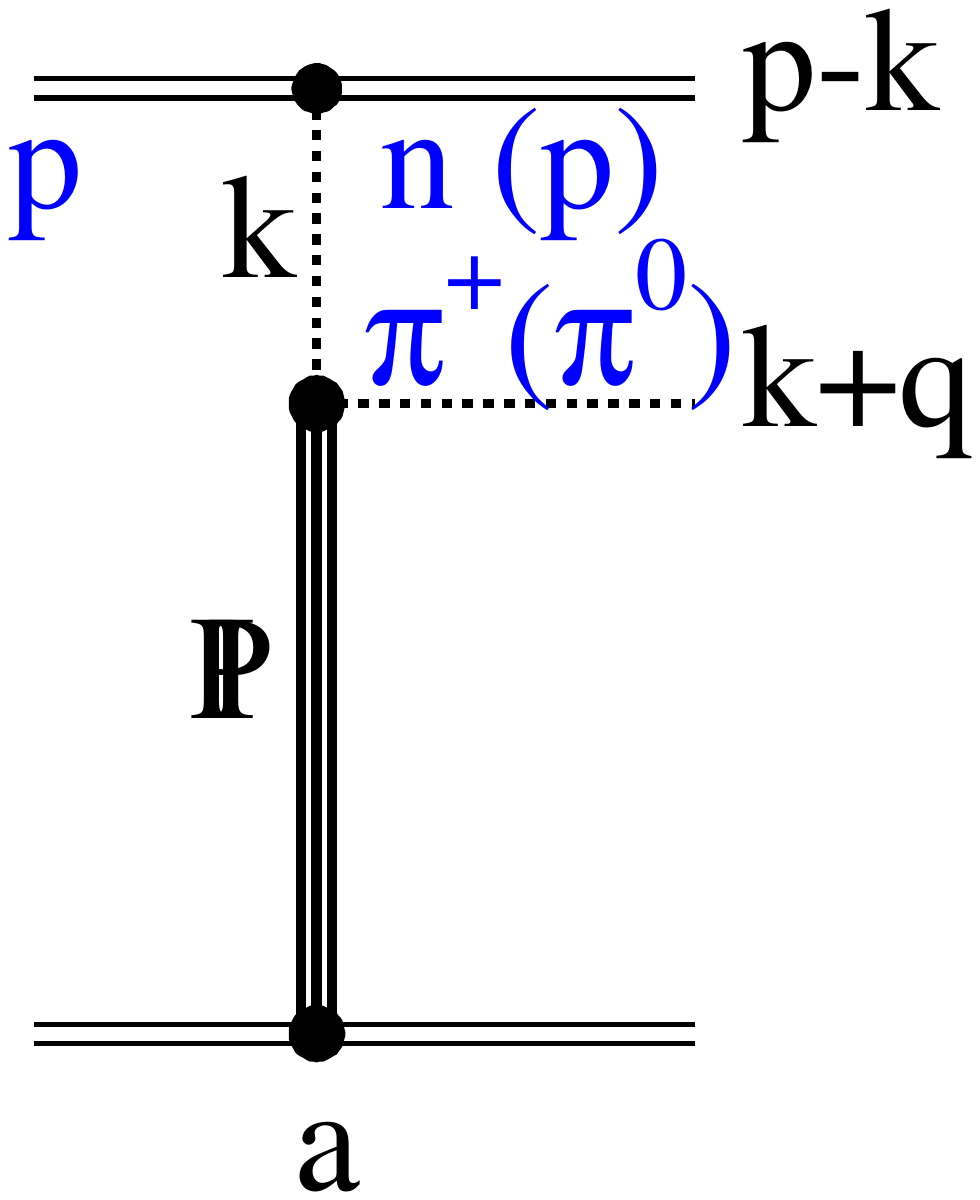}
\hspace{-3.9cm}
\includegraphics[scale=0.33]{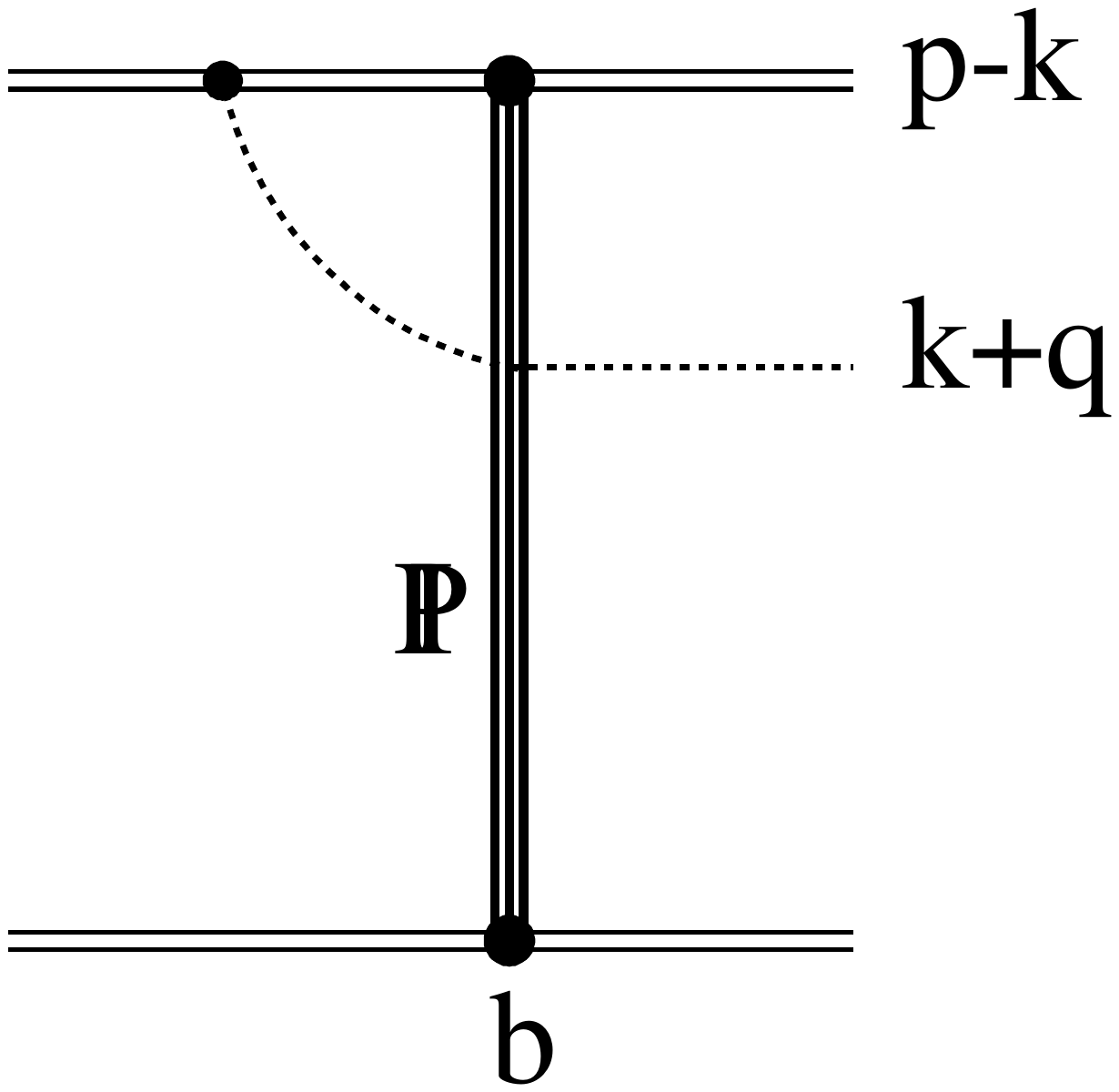}
\hspace{-4.5cm}
\includegraphics[scale=0.33]{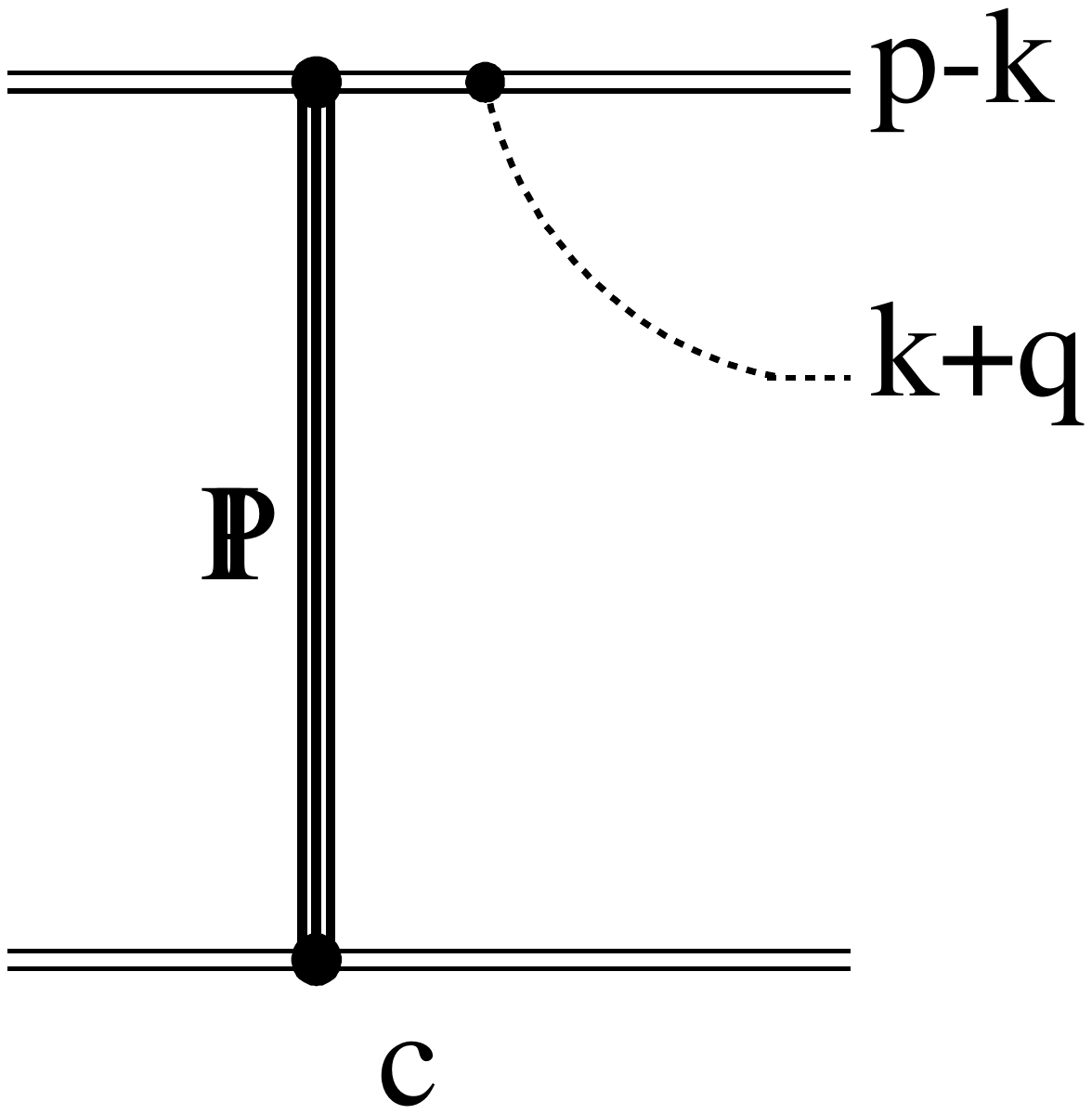}
\vspace{-2cm}
\caption{\sf The nuclear 'Deck' amplitudes for low-mass proton dissociation.}
\label{fig:3}
\end{figure}

  First, the amplitude shown in Fig.3a reads
$$A^{3a}=G_{\pi N}G(k^2)~\frac{\sqrt{-k^2}}{k^2-m^2_\pi}~ T_{\pi p}(xs,t)$$
\begin{equation}
\label{3a}
=~G_{\pi N}G(k^2)~~\frac{\sqrt{-k^2}}{k^2-m^2_\pi} ~~ i(1-i\rho)xs \frac 23\sigma_0\left(\frac{xs}{s_0}\right)^{\alpha_P(t)-1}F_{\pi}(t)F_{p}(t)   \ ,
\end{equation}
where $x$ is the beam momentum fraction carried by the pion, $s$ is the initial energy
{\cred {squared}} and $T_{\pi p}$ is the amplitude of the strong $\pi p$ interaction parametrized in the second line of the above equation by the pomeron pole exchange with effective trajectory $\alpha_P(t)=1+\Delta+\alpha'_Pt$, with Im$T_{p p}(s)= s\sigma_0(s/s_0)^{\alpha_P(0)-1}$, in which we take $\Delta=0.1$ and $\alpha'_P=0.25$ GeV$^2$. As usual $s_0=1$ GeV$^2$. The coupling $G_{\pi N}=G_{\pi N}(k^2=0)$ for the $\gamma_5$ proton pion vertex\footnote{For $\pi^+$ the coupling is $\sqrt{2}$ larger than for $\pi^0$.} is $G^2_{\pi^0pp}/4\pi=13.75$~\cite{G-N-pi} at $k^2=0$ with the dipole form factor 
\be
G(k^2)=1/(1-k^2/0.71{\rm GeV}^2)^2\ .
\ee
$m_\pi$ is the pion mass, and we will take $m$ to be the mass of the proton.
    
Besides the contribution from the term {\cred {$\alpha'_Pt=\alpha'_Pq^2$}}, the $q^2=t$ dependence of
{\cred {the}} strong amplitude is driven by the `form factors' in 
{\cred {the}} vertices 
\be
F_{p}(q^2)=1/(1-q^2/0.71{\rm GeV}^2)^2,
\ee
\be
 F_{\pi}(q^2)=1/(1-q^2/0.6{\rm GeV}^2).
 \ee

Analogously, the amplitudes corresponding to {\cred {Figs.3b,c}} are
\begin{equation}
\label{3b}
A^{3b}~=~G_{\pi N}G(k^2)~~\frac{\sqrt{-k^2}}{(p-k-q)^2-m^2}~~i(1-i\rho)(1-x)s\sigma_0\left(\frac{(1-x)s}{s_0}\right)^{\alpha_P(q^2)-1}F^2_{p}(q^2)\ ,
\end{equation}
and
\begin{equation}
\label{3c}
A^{3c}=G_{\pi N}G(k^2)~~\frac{\sqrt{-k^2}}{(p+q)^2-m^2} ~~i(1-i\rho)s \sigma_0 \left(\frac s{s_0}\right)^{\alpha_P(q^2) -1}F^2_{p}(q^2)\ .
\end{equation}

{\cred {For completeness}} we give the formulae for the propagators:
\begin{equation}
k^2-m^2_\pi=-\frac 1{1-x}(k^2_t+x^2m^2)-m^2_\pi,\;\;\;~~~ (p+q)^2-m^2=\Delta M^2
\end{equation}
$$ (p-k-q)^2-m^2=m^2_\pi-k^2-q^2_t-\Delta M^2\ ,$$
where
\be
\Delta M^2=(m^2+q^2_t)/(1-x)+(m^2_\pi+(k+q)^2_t)/x-m^2.
\ee
Note that here the values of $k^2_t$, $q^2_t$ and $(k+q)^2_t$ are positive.
At very high energies $s\gg q_t^2$ where the photon virtuality $q^2=-q^2_t$.

\begin{figure}[htb]
\vspace{-3cm}
\hspace{2cm}
\includegraphics[scale=0.33]{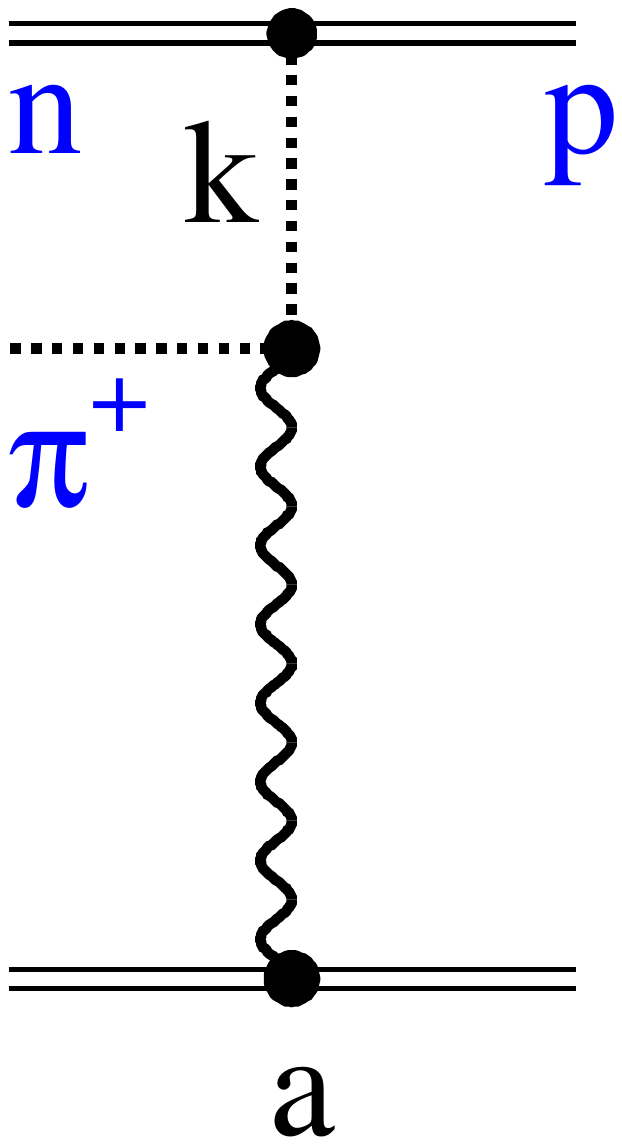}
\hspace{-3.8cm}
\includegraphics[scale=0.33]{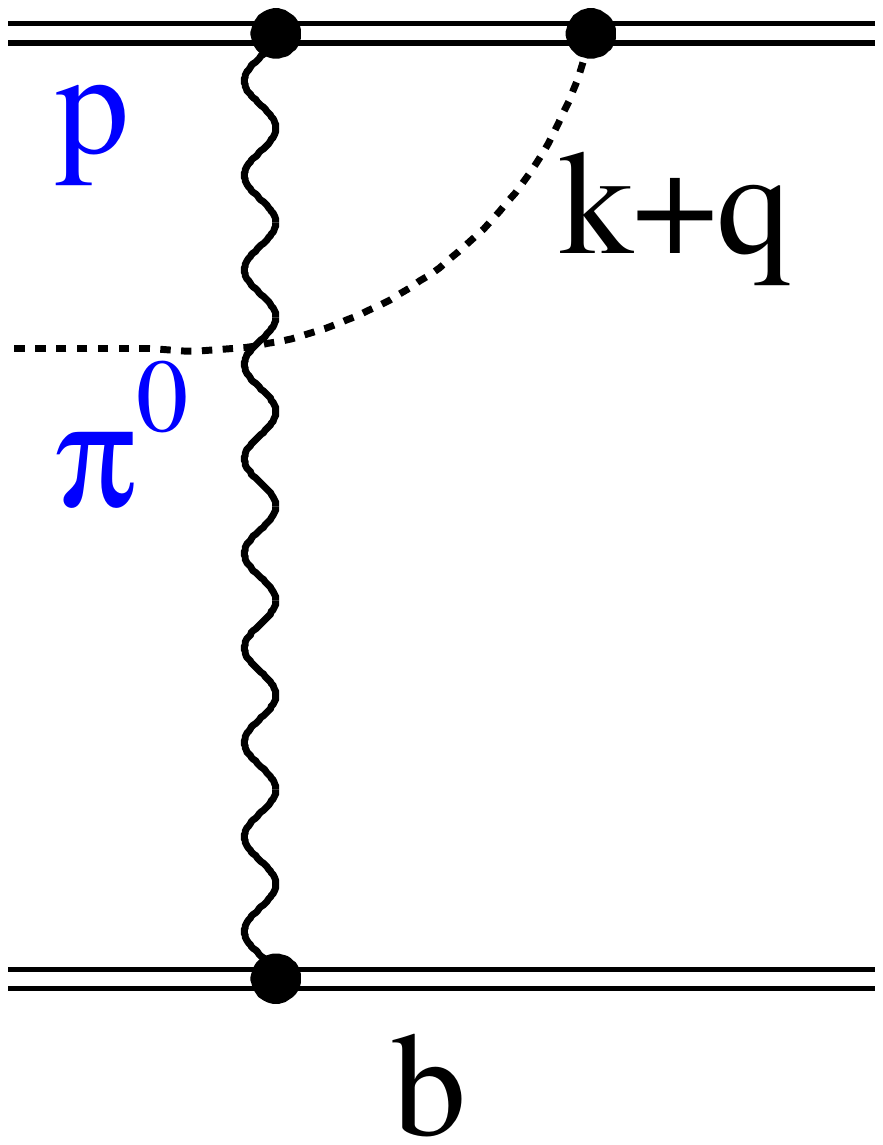}
\hspace{-4.9cm}
\includegraphics[scale=0.33]{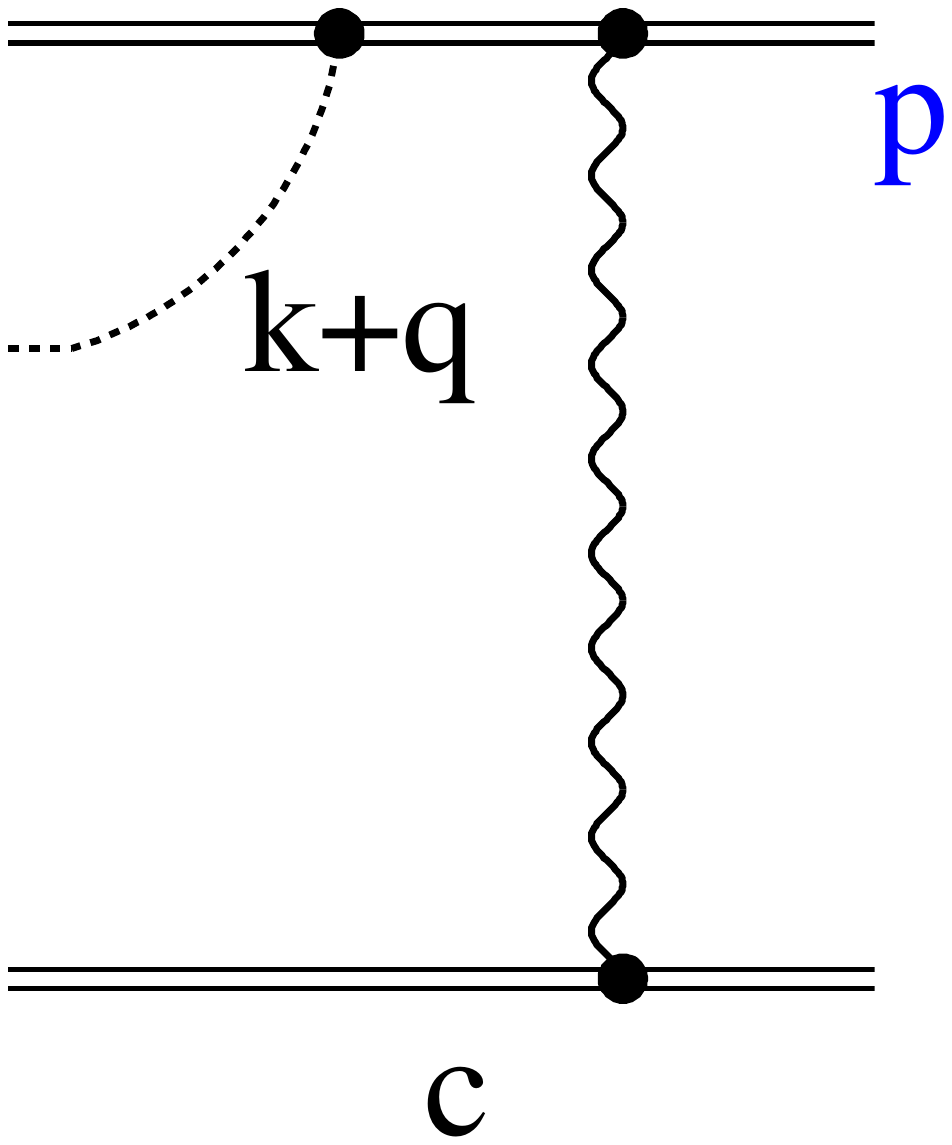}
\vspace{-1.5cm}
\caption{\sf The QED `Deck' amplitudes for the low-mass proton dissociation.}
\label{fig:4}
\end{figure}

The QED amplitudes of Fig.4 take the form
\begin{equation}
\label{4a}
A^{4a}~=~G_{\pi N}G(k^2) ~~\frac {\sqrt{-k^2}}{k^2-m^2_\pi}~~\frac{8\pi\alpha}{q^2}xsF_{\pi}(q^2)F_{p}(q^2)\ ,
\end{equation}
\begin{equation}
\label{4b}
A^{4b}~=~G_{\pi N}G(k^2) ~~\frac {\sqrt{-k^2}}{(p-k-q)^2-m^2}~~\frac{8\pi\alpha}{q^2}(1-x)sF^2_{p}(q^2)\ ,
\end{equation}
and
\begin{equation}
\label{4c}
A^{4c}~=~G_{\pi N}G(k^2) ~~ \frac {\sqrt{-k^2}}{(p+q)^2-m^2} ~~ \frac{8\pi\alpha}{q^2}sF^2_{p}(q^2))\ .
\end{equation}
Again a dipole form factor 
\be
F_p=F_{p\gamma}(q^2)=1/(1-q^2/0.71{\rm GeV}^2)^2
\ee
 is used for the photon-proton vertices
while for the pion-photon coupling we take the pole form 
\be
F_\pi=F_{\pi\gamma}(q^2)=1/(1-q^2/0.6{\rm GeV}^2).
\ee
Recall that for the case of $\pi^+$ the coupling $G_{\pi N}$ must be multiplied by $\sqrt 2$ and  we have to deal with the sum $A^{4a}+A^{4c}$ while the {\cred{total}} QED  amplitude with a
$\pi^0$ meson is given by $A^{4b}+A^{4c}$.
It is easy to check that {\cred{the total}} QED amplitude of proton excitation vanishes as $q_t\to 0$.

The product of {\cred{the total}}  `nuclear' times the {\cred {total}} QED amplitudes now has to be integrated over the momentum fraction $x$ and the transverse momenta $q_t$ and $k_t$. Recall that we are 
{\cred {seeking}} for the phase $\phi$ at $t=0$. We find
\begin{equation}
\label{a-phi}
\alpha\phi^{\rm Deck}=\frac 2{32\pi^2s^2\sigma(pp)}\int_0^1 \frac{dx}{x(1-x)}\int dq^2_t\int d^2k_t ({\rm Im}A^{(3)})\cdot A^{(4)}\ , 
\end{equation}
where $A^{(3)}$ and $A^{(4)}$ denote the {\cred{total}} amplitudes, that is the sum of the corresponding $a,b,c$ contributions. The factor 2 accounts for the dissociation of the lower proton. The denominator in $dx/(x(1-x))$
{\cred {arises}} from the $1/(2E_\pi\ 2E_N)$ factors in {\cred{the}} phase space integrals $d^3k/(2E(2\pi)^3)$.

Numerical calculation at $\sqrt s=13$ TeV results in $\alpha\phi^{\rm Deck}=1.3\cdot 10^{-4}$, which is negligibly small  in comparison with the experimental accuracy of  0.01. In terms of the Bethe phase, the `inelastic' diagrams with proton low-mass excitations change $\phi$ by about 0.018~\footnote{The reason for such a small contribution from proton dissociation is {\cred{as follows}}. The low-mass nucleon photo-excitation is mainly {\cred{a}} magnetic transition which flips the proton helicity. Indeed, the spin flip in $N\to N^*$ transition
 is needed in order to compensate for the spin=1 of $\gamma$
{\cred {quantum}} in {\cred {the}} $N^*\to p\gamma$ decay. On the other hand, the pomeron exchange amplitude contains two components: one conserving the $s$-channel helicity and another one which flips the helicity. The second component acts as the anomalous magnetic moment.
Let us assume that the Pomeron-nucleon vertex is similar to the photon-nuclear vertex~\cite{SHC}. Then the term responsible for the spin flip component is given by the anomalous magnetic moment for zero isospin ($I=0$) exchange amplitude. That is 
 for the diffractive 
  transition 
  $\mu_{I=0}=(\mu_p+\mu_n)/2=(1.79-1.91)/2=0.06$ is very small.}. A similar variation (0.02) of $\phi$ was observed in \cite{Cahn} depending on the form of the parametrization of the `elastic' proton form factor -(that is either dipole or exponent).

\subsection{Deck cross section}
The cross section of low-mass dissociation given by the Deck diagram {\cred{ shown in}} Fig.2a reads
\begin{equation}
\label{D-sig}
\sigma^{\rm SD}~=~\frac 1{4s^2\cdot (4\pi)^3}\int_0^1\frac{dx}{x(1-x)}\int dq^2_t\int dk^2_t  A^{(3)}\cdot A^{*(3)}\ .
\end{equation}
Here we account for the dissociation of only one of the colliding proton. For $\sqrt s=7$ TeV we note that the total cross section\footnote{This is the value between the cross sections given by TOTEM~\cite{TOT-tot} and by ATLAS-ALFA~\cite{ALFA-tot}.} $\sigma_{\rm tot}=97$ mb and ({\ref{D-sig}) gives $\sigma^{\rm SD}=1.35$ mb. This is to be compared with low-mass dissociation cross section $2.6/2=1.3\pm 1.1$ mb observed by TOTEM \cite{TOTEM2}. The good agreement confirms the applicability of our calculation of 
{\cred {the}} low-mass proton {\cred {excitation}} contribution to {\cred {the}} Bethe phase  $\phi$.

\section{Higher-mass dissociation}
To evaluate the possible role of higher-mass excitations we consider the `triple-Regge'-like diagram of Fig.2b. Since the $RI\!\!P\gamma$ triple vertex is not known phenomenologically we use the `Vector Dominance Model' (VDM)~\cite{VDM} approach and replace {\cred {this vertex}} by the 
`Pomeron -- $\rho$-meson' (or $\omega$-meson) vertex which in its turn {\cred {can}} be written as 2/3 of the Pomeron-proton vertex. 
 Recall that due to gauge invariance  the proton excitation vertex caused by the 
photon must {\cred {vanish}} as $q^2\to 0$.  
The dimension of the corresponding $q^2$ factor should be compensated either by the radius of {\cred {the}} $RI\!\!P\gamma$ triple vertex or by the mass difference $\Delta M^2=M^2-m^2$. In the present calculation we use $s_0=1$ GeV$^2$. On the one hand, this {\cred {simplifies}} the final Regge formula, while on {\cred{the other}} hand this is close to the expected size of the vertex driven by the slope of the $R$-reggeon ($\rho,\ \omega$) trajectory $\alpha'_R=0.9$ GeV$^{-2}$~\cite{Irv}.

Next, it is known within the VDM, that the proton-to-photon coupling (proton electric charge $e$) can 
be {\cred {considered}} as the sum of the contributions mediated by {\cred{the}}
$\rho$ and $\omega$ mesons. Exploiting the fact that the $\rho$ and $\omega$ Regge trajectories are {\cred {degenerate}}~\cite{Irv}
 we calculate the contribution {\cred {shown in Fig.2b}} as
\begin{equation}
\label{eq:2b}
\alpha\phi^{R}=\alpha'_R\pi\frac 23\frac \alpha\pi\int \frac{dM^2_X}{M^2_X}\left(\frac{M^2_X}{s_0}\right)^{\alpha_R(0)-\alpha_P(0)}\int dq^2_t
F^2_{p}(q^2)\left(\frac{M^2_X}s\right)^{\alpha'_Pq^2}\ .
\end{equation} 
Here $M_X$ is the mass of the proton- {\cred {excited}} system described by the $R$-reggeon and we {\cred {have}} already accounted for the possibility of excitation of the lower proton. 

The first factor $\alpha'_R\pi$ in (\ref{eq:2b}) accounts for the relation between the imaginary part of the reggeon exchange amplitude given by  the $R$-reggeon signature factor
\begin{equation}
\label{signat}
\eta~=~\frac{1-\exp(-i\pi\alpha_R(t))}{\sin(-\pi\alpha_R(t))}
\end{equation}
and the residue of the pole at $\alpha_R(t)=1$. Near the pole the signature
{\cred{factor}}
 (\ref{signat}) takes the form $2/(\alpha'_R\pi(t-m^2_R))$ while 
 the discontinuity at $t=0$ (where $\alpha_R(0)\simeq 1/2$) is $2{\rm Im}\eta\simeq 2$.
 
 The numerical calculation of (\ref{eq:2b}) results in
 \begin{equation}
 \label{phi-2b}
 \phi^R~=~0.099\ -\ 0.106\ 
 \end{equation}
 for $\alpha_R(0)=0.5~-~0.54$.
{\cred{This leads to}} a correction
\be
\alpha\phi^R\simeq 0.0007~-~0.0008
\ee
to the $\rho$=Re/Im ratio for the `nuclear' amplitude.

Recall that for this evaluation we  used a very approximate approach. Nevertheless, the result is an order-of-magnitude less than the accuracy of the present experiment (see~\cite{TOTEM}). Moreover, most probably the true value of $\phi^R$ is even smaller since, as a rule, the triple-Reggeon vertices extracted from the phenomenological triple-Regge analysis are smaller
than the corresponding Reggeon-hadron vertices (see for example~\cite{Luna}, and the discussion below).

In terms of VDM the interaction with the photon starts from the transition of a point-like photon to the $q\bar q$ pair where the quark-quark separation, $r$, is close to zero. On the other hand the Pomeron-induced cross section of such a pair behaves as $\sigma\propto \alpha^2_s\langle r^2\rangle $~\cite{Br-Kop}. Most probably the time interval occupied by the $RI\!\!P\gamma$ interaction is not sufficient to form the $\rho-$ or $\omega-$ mesons in their normal (equilibrium) states, so the resulting values of $r^2$, which drive the value of the vertex, will be smaller than that used in our estimate 
(based on the assumption that $\sigma_{pR}=\sigma_{\omega p}=(2/3)\sigma_{pp}$).
 This qualitatively explains why we expect that the value of $\phi^R$ to be smaller than that calculated above.

Note also that strictly speaking one should not sum the phases $\phi^{\rm Deck}$ and $\phi^R$. This will lead to double counting since {\cred{when}} calculating $\phi^R$ {\cred{using}}  (\ref{eq:2b}) we
{\cred{integrate}} over $M_X$ starting from $M_X=s_0=1$ GeV$^2$. If {\cred {we}} would like to keep the contribution described by {\cred{the}} Deck diagrams then in (\ref{eq:2b}) we {\cred{have}} to take a larger lower limit for $M_X$. This will diminish the value of $\phi^R$.

\section{ Two-photon exchange with proton excitation}

The `inelastic' contribution of the two-photon exchange diagram 
shown in Fig.1c 
 can be calculated using the equivalent photon approximation~\cite{WW}. 
The imaginary part of {\cred{the}} amplitude {\cred {in Fig.1c}} reads
\begin{equation}
\label{eq:1c}
A^{1c}=2\frac\alpha{\pi^2}s\int\frac{dE_\gamma}{E_\gamma}\int d^2q_t\frac{(q_1\cdot q_2)}
{q^2_1q^2_2}\sigma^{\rm tot}_{\gamma p}(E_\gamma)F_{p}(q^2_1)F_{p}(q^2_2)\ ,
\end{equation}
where first factor 2 accounts for the excitations of the second (lower in Fig.1c) proton.
Here we have to be more precise and to account for the small but non-zero {\cred{total}} momentum transferred $t=Q^2=-Q^2_t$. The momenta of the `left' and the `right' photons in Fig.1c are 
\begin{equation}
q_{1,2}=q_t\pm \frac{Q_t}2
\end{equation}
and $(q_1\cdot q_2)$ denotes the scalar product of $q_1$ and $q_2$.\; $E_\gamma$ is the photon energy in the upper (in Fig.1c) proton  rest frame; $\sigma^{\rm tot}_{\gamma p}$ is the total cross section of photon-proton interaction. 

The resulting value of $A^{1c}$ in (\ref{eq:1c}) should be compared with the one-photon exchange (Coulomb) amplitude (which is real)
\begin{equation}
\label{C}
F^C(t)=s\frac{8\pi\alpha}{Q^2}\ .
\end{equation} 
Note that, {\cred{contrary}} to $F^C$, the proton excitation contribution $A^{1c}$ of (\ref{eq:1c}) does not contain a $1/Q^2$ pole. Therefore, the phase {\cred {generated}} by the
$A^{1c}/F^C$ ratio vanishes at $t=Q^2\to 0$. However actually the Coulomb-nuclear interference is measured at {\cred {$|t|\sim 0.001\ {\rm GeV}^2\neq 0$.}} That is why we wrote the formula (\ref{eq:1c}) accounting for the value of $Q_t$.

For the numerical estimate we take the experimentally measured $\sigma^{\rm tot}_{\gamma p}(E_\gamma)$ cross sections~\cite{PRD,PRC} at $E_\gamma=0.26~-~4.2$ GeV.  For a larger $E_\gamma>4$ GeV we {\cred {use}} parametrization of~\cite{PRD} \be
\sigma^{\rm tot}_{\gamma p}(E_\gamma)=(91+71.4/\sqrt{E_\gamma})\mu{\rm b}
\ee
with $E_\gamma$ in GeV. In this parametrization we keep only the second term since the first term corresponds to Pomeron exchange ($\sigma={\rm const})$ and should be treated as an  $O(\alpha^2)$ correction to the strong interaction (even-signature) amplitude.

 \begin{figure}[htb]
 \vspace{-5cm}
 \hspace{3.5cm}
\includegraphics[scale=0.5]{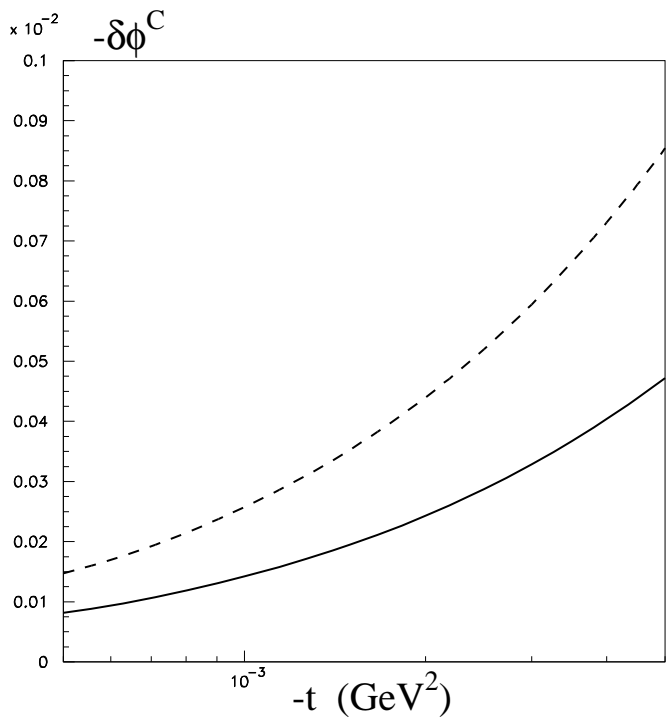}
\caption{\sf The phase shift $\delta\phi^C$ of the one-photon-exchange amplitude caused by the second photon exchange with proton excitations in 
{\cred {the}} intermediate states. The dashed line is calculated using the full photon-proton cross section, $\sigma^{\rm tot}_{\gamma P}(E_\gamma)$ at $E_\gamma<4.2$ GeV, while for the solid curve the Pomeron (constant) "background" of 91 $\mu$b was subtracted from $\sigma^{\rm tot}_{\gamma p}$.}.
\label{fig:c}
\end{figure}
As seen from Fig.5, in the region of interest ($|t|<0.001-0.005$ GeV$^2$), where Coulomb-nuclear interference {\cred{manifests}} itself,
 the possibility of proton excitations 
{\color{black}  in the two-photon exchange process} changes the original phase of 
{\cred {the}} pure QED one-photon-exchange amplitude by {\cred{the}} negligibly small value of $|\delta\phi^C|<10^{-3}$.

\section{Conclusion}
We evaluated the contribution of proton ($p\to N^*$) excitations to the phase shift (Bethe phase) between the strong interaction and the one-photon exchange QED amplitudes caused by an additional photon exchange. The low-mass part was calculated based} on the Deck~\cite{Deck} ($p\to N\pi$) {\cred {mechanism}}, while the higher-mass  {\cred {excitation}} was estimated using the triple-Regge formalism. The `inelastic' two-photon exchange QED contribution was calculated using the experimental data on {\cred {the}} proton-photon cross section in terms of the equivalent photon approach.

It {\cred {is}} shown that the effects are very small and do not change the value of  $\rho={\rm Re/Im}$ ratio, measured via the Coulomb-nuclear interference in small angle elastic $pp$ scattering, {\cred {by}} more than $\delta\rho\sim 10^{-3}$. This is about {\cred {an}} order-of-magnitude less than the experimental accuracy of $\pm 0.01$~\cite{TOTEM}.

\section{Acknowledgements}
 We thank Per Grafstrom for the interest to our work and  useful comments. 
{\cred{ MGR thanks the IPPP at the University of Durham for hospitality.}}  

\end{document}